# Meaning of the wave function

Shan Gao

Unit for HPS & Centre for Time, SOPHI, University of Sydney

Email: sgao7319@uni.sydney.edu.au

We investigate the meaning of the wave function by analyzing the mass and charge density distributions of a quantum system. According to protective measurement, a charged quantum system has effective mass and charge density distributing in space, proportional to the square of the absolute value of its wave function. In a realistic interpretation, the wave function of a quantum system can be taken as a description of either a physical field or the ergodic motion of a particle. If the wave function is a physical field, then the mass and charge density will be distributed throughout space at a given time for a charged quantum system, and thus there will exist gravitational and electrostatic self-interactions of its wave function. This not only violates the superposition principle of quantum mechanics but also contradicts experimental observations. Thus the wave function cannot be a description of a physical field but a description of the ergodic motion of a particle. For the later there is only a localized particle with mass and charge at every instant, and thus there will not exist any self-interaction for the wave function. It is further argued that the classical ergodic models, which assume continuous motion of particles, cannot be consistent with quantum mechanics. Based on the negative result, we suggest that the wave function is a description of the quantum motion of particles, which is random and discontinuous in nature. On this interpretation, the square of the absolute value of the wave function not only gives the density of probability of the particle being found in certain locations, but also gives the density of objective probability of the particle being there. We show that this new interpretation of the wave function provides a natural realistic alternative to the orthodox interpretation, and its implications for other realistic interpretations of quantum mechanics are also briefly discussed.

## 1. Introduction

The wave function is the most fundamental concept of quantum mechanics. It was first introduced into the theory by analogy (Schrödinger 1926); the behavior of microscopic particles likes wave, and thus a wave function is used to describe them. Schrödinger originally regarded the wave function as a description of real physical wave. But this view met serious objections and was soon replaced by Born's probability interpretation (Born 1926), which becomes the standard interpretation of the wave function today. According to this interpretation, the wave function is a probability amplitude, and the square of its absolute value represents the probability density for a particle to be measured in certain locations. However, the standard interpretation is still unsatisfying when applying to a fundamental theory because of resorting to measurement (see, e.g. Bell 1990). In view of this problem, some alternative realistic interpretations of the wave function have been proposed and widely studied (Bohm 1952; Everett 1957; Nelson 1966; Ghirardi, Grassi and Benatti 1995).

There are in general two possible ways to interpret the wave function of a single quantum system in a realistic interpretation[1]. One view is to take the wave function as a physical entity simultaneously distributing in space such as a field, and it is assumed by de Broglie-Bohm theory, many-worlds interpretation and dynamical collapse theories etc (de Broglie 1928; Bohm 1952; Everett 1957; Ghirardi, Grassi and Benatti 1995)[2]. For example, in de Broglie-Bohm theory the wave function is generally taken as an objective physical field, called $\Psi$-field, though there are various views on exactly what field the wave function is. The other view is to take the wave function as a description of some kind of ergodic motion of a particle (or corpuscle), and it is assumed by stochastic interpretation etc. The essential difference between a field and the ergodic motion of a particle lies in the property of simultaneity. The field exists throughout space simultaneously, whereas the ergodic motion of a particle exists throughout space in an essentially local way; the particle is still in one position at each instant, and it is only during a time interval that the ergodic motion of the particle spreads throughout space.

It is widely expected that the correct realistic interpretation of the wave function can only be determined by future precise experiments. In this paper, we will argue that the above two interpretations of the wave function can in fact be tested by analyzing the mass and charge density distributions of a quantum system, and the former has already been excluded by experimental observations. Moreover, a further analysis can also determine which kind of ergodic motion of particles the wave function describes. The plan of this paper is as follows. In Section 2, we first argue that a quantum system with mass $m$ and charge $Q$, which is described by the wave function $\psi(x,t)$, has effective mass and charge density distributions $m|\psi(x,t)|^2$ and $Q|\psi(x,t)|^2$ in space respectively. This argument is strengthened in Section 3 by showing that the result is also a consequence of protective measurement. In Section 4, we argue that the field explanation of the wave function entails the existence of an electrostatic self-interaction for the wave function of a charged quantum system, as the charge density will be distributed in space simultaneously for a physical field. This contradicts the predictions of quantum mechanics as well as experimental observations. Thus we conclude that the wave function cannot be a description of a physical field. This leads us to the second view that interprets the wave function as a description of the ergodic motion of particles. In Section 5, it is argued that the classical ergodic models, which assume continuous motion of particles, cannot be consistent with quantum mechanics, and thus they have been excluded. Section 6 further investigates the possibility that the wave function is a description of the quantum motion of particles, which is random and discontinuous in nature. It is shown that this new interpretation of the wave function provides a natural realistic alternative to the orthodox interpretation, and its implications for other realistic interpretations of quantum mechanics are also briefly discussed.

---

[1] For the sake of simplicity, we will mainly discuss the wave function of a single quantum system in this paper. The conclusion can be readily extended to many-body system, which wave function is defined in configuration space.
[2] Note that the wave function in de Broglie-Bohm theory is also regarded as nomological, e.g. a component of physical law rather than of the reality described by the law (Dürr, Goldstein and Zanghì 1997; Goldstein and Teufel 2001). We will not discuss this view in this paper. But it might be worth noting that this non-field view may have serious drawbacks when considering the contingency of the wave function (see, e.g. Valentini 2009), and the results obtained in this paper seemingly disfavor this view too. Besides, we note that the field interpretation may be debatable for many-worlds interpretation. But according to Everett (1957), "the wave function is taken as the basic physical entity with no a priori interpretation", and "observers and object systems…They all are represented in a single structure, the field".

## 2. How do mass and charge distribute for a single quantum system?

The mass and charge of a charged classical system always localize in a definite position in space at each moment. For a charged quantum system described by the wave function $\psi(x,t)$, how do its mass and charge distribute in space then? We can measure the total mass and charge of the quantum system by gravitational and electromagnetic interactions and find them in some region of space. Thus the mass and charge of a quantum system must also exist in space with a certain distributions if assuming a realistic view. Although the mass and charge distributions of a single quantum system seem meaningless according to the probability interpretation of the wave function, it should have a physical meaning in a realistic interpretation of the wave function such as de Broglie-Bohm theory[3].

As we think, the Schrödinger equation of a charged quantum system under an external electromagnetic potential already provides an important clue. The equation is

$$i\hbar \frac{\partial \psi(x,t)}{\partial t} = \left[ -\frac{\hbar^2}{2m}\left(\nabla - \frac{iQ}{\hbar c}A\right)^2 + Q\varphi \right]\psi(x,t) \qquad (1)$$

where $m$ and $Q$ is respectively the mass and charge of the system, $\varphi$ and $A$ are the electromagnetic potential, $\hbar$ is Planck's constant divided by $2\pi$, $c$ is the speed of light. The electrostatic interaction term $Q\varphi\psi(x,t)$ in the equation seems to indicate that the charge of the quantum system distributes throughout the whole region where its wave function $\psi(x,t)$ is not zero. If the charge does not distribute in some regions where the wave function is nonzero, then there will not exist any electrostatic interaction there. But the term $Q\varphi\psi(x,t)$ implies that there exists an electrostatic interaction in all regions where the wave function is nonzero. Thus it seems that the charge of the quantum system should distribute throughout the whole region where its wave function is not zero. Furthermore, since the integral $\int_{-\infty}^{+\infty} Q|\psi(x,t)|^2 dx$ is the total charge of the system, the charge density distribution in space will be $Q|\psi(x,t)|^2$. Similarly, the mass density can be obtained from the Schrödinger equation of a quantum system with mass $m$ under an external gravitational potential $V_G$:

---

[3] Unfortunately it seems that the orthodox probability interpretation of the wave function still influences people's mind even if they already accept a realistic interpretation of the wave function. One obvious example is that few people admit that the realistic wave function has energy density (Holland (1993) is a notable exception). If the wave function has no energy, then it seems very difficult to regard it as physically real. Even if Bohm and Hiley (1993) interpreted the Ψ-field as "active information", they also admitted that the field has energy, though very little. Once one admits that the wave function has energy density, then it seems natural to endow it with mass and charge density, which are two common sources of energy density.

$$i\hbar \frac{\partial \psi(x,t)}{\partial t} = \left[ -\frac{\hbar^2}{2m}\nabla^2 + mV_G \right] \psi(x,t) \tag{2}$$

The gravitational interaction term $mV_G\psi(x,t)$ in the equation also indicates that the (passive gravitational) mass of the quantum system distributes throughout the whole region where its wave function $\psi(x,t)$ is not zero, and the mass density distribution in space is $m|\psi(x,t)|^2$.

The above result can be more readily understood when the wave function is a complete realistic description of a single quantum system as in many-worlds interpretation and dynamical collapse theories. If the mass and charge of a quantum system does not distribute as above in terms of its wave function $\psi(x,t)$, then other supplement quantities will be needed to describe the mass and charge distributions of the system in space, while this obviously contradicts the premise that the wave function is a complete description. In fact, the dynamical collapse theories such as GRW theory already admit the existence of mass density (Ghirardi, Grassi and Benatti 1995).

In addition, even in de Broglie-Bohm theory, which takes the wave function as an incomplete description and admits supplement hidden variables (i.e. the trajectories of Bohmian particles accompanying the wave function), there are also some arguments for the above mass and charge density explanation (Holland 1993; Brown, Dewdney and Horton 1995). It was argued that since the Ψ-field depends on the parameters such as mass and charge, it may be said to be massive and charged (Holland 1993). Brown, Dewdney and Horton (1995), by examining a series of effects in neutron interferometry, argued that properties sometimes attributed to the "particle" aspect of a neutron, e.g., mass and magnetic moment, cannot straightforwardly be regarded as localized at the hypothetical position of the particle in Bohm's theory. They also argued that it is hard to understand how the Aharonov-Bohm effect is possible if that the charge of the electron which couples with the electromagnetic vector-potential is not co-present in the regions on all sides of the confined magnetic field accessible to the electron (Brown, Dewdney and Horton 1995).

One may object that de Broglie-Bohm theory and many-worlds interpretation seemingly never admit the above mass density explanation, and no existing interpretation of quantum mechanics including dynamical collapse theories endows charge density to the wave function either. As we think, however, protective measurement provides a more convincing argument for the existence of mass and charge density distributions[4]. The wave function of a single quantum system, especially its mass and charge density, can be directly measured by protective measurement. Therefore, a realistic interpretation of quantum mechanics should admit the mass and charge density explanation in some way; if it cannot, then it will be at least problematic concerning its explanation of the wave function.

---

[4] It is very strange for the author that most supporters of a realistic interpretation of quantum mechanics ignore protective measurement and its implications. Admittedly there have been some controversies about the meaning of protective measurement (see, e.g. see e.g. Rovelli 1994; Uffink 1999; Dass and Qureshi 1999), but the debate mainly centers on the reality of the wave function. If one insists on a realistic interpretation of quantum mechanics such as de Broglie-Bohm theory, then the debate will be mostly irrelevant and protective measurement will have strict restrictions on the realistic interpretation.

## 3. Protective measurement and its answer

In this section, we will give a brief introduction of protective measurement and its implication for the existence of mass and charge density distributions. Different from the conventional measurement, protective measurement aims at measuring the wave function of a single quantum system by repeated measurements that do not destroy its state. The general method is to let the measured system be in a non-degenerate eigenstate of the whole Hamiltonian using a suitable interaction, and then make the measurement adiabatically so that the wave function of the system neither changes nor becomes entangled with the measuring device appreciably. The suitable interaction is called the protection.

As a typical example of protective measurement (Aharonov, Anandan and Vaidman 1993; Aharonov, Anandan and Vaidman 1996), we consider a quantum system in a discrete nondegenerate energy eigenstate $\psi(x)$. The protection is natural for this situation, and no additional protective interaction is needed. The interaction Hamiltonian for measuring the value of an observable $A$ in the state is:

$$H_I = g(t)PA \tag{3}$$

where $P$ denotes the momentum of the pointer of the measuring device, which initial state is taken to be a Gaussian wave packet centered around zero. The time-dependent coupling $g(t)$ is normalized to $\int_0^T g(t)dt = 1$, where $T$ is the total measuring time. In conventional von Neumann measurements, the interaction $H_I$ is of short duration and so strong that it dominates the rest of the Hamiltonian (i.e. the effect of the free Hamiltonians of the measuring device and the system can be neglected). As a result, the time evolution $\exp(-iPA/\hbar)$ will lead to an entangled state: eigenstates of $A$ with eigenvalues $a_i$ are entangled with measuring device states in which the pointer is shifted by these values $a_i$. Due to the collapse of the wave function, the measurement result can only be one of the eigenvalues of observable $A$, say $a_i$, with a certain probability $p_i$. The expectation value of $A$ is then obtained as the statistical average of eigenvalues for an ensemble of identical systems, namely $<A> = \sum_i p_i a_i$. By contrast, protective measurements are extremely slow measurements. We let $g(t) = 1/T$ for most of the time $T$ and assume that $g(t)$ goes to zero gradually before and after the period $T$. In the limit $T \to \infty$, we can obtain an adiabatic process in which the system cannot make a transition from one energy eigenstate to another, and the interaction Hamiltonian does not change the energy eigenstate. As a result, the corresponding time evolution $\exp(-iP<A>/\hbar)$ shifts the pointer by the expectation value $<A>$. This result strongly contrasts with the conventional measurement in which the pointer shifts by one of the eigenvalues of *A*.

It should be stressed that $T \to \infty$ is only an ideal situation[5], and a protective measurement can never be performed on a single quantum system with absolute certainty because of the tiny unavoidable entanglement (see also Dass and Qureshi 1999)[6]. For example, for any given values of *P* and *T*, the energy shift of the above eigenstate, given by first-order perturbation theory, is

$$\delta E = <H_I> = \frac{<A>P}{T} \qquad (4)$$

Correspondingly, we can only obtain the exact expectation value $<A>$ with a probability very close to one, and the measurement result can also be the expectation value $<A>_\perp$, with a probability proportional to $1/T^2$, where $\perp$ refers to the normalized state in the subspace normal to the initial state $\psi(x)$ as picked out by first-order perturbation theory (Dass and Qureshi 1999). Therefore, an ensemble, which may be considerably small, is still needed for protective measurements.

Although a protective measurement can never be performed on a single quantum system with absolute certainty, the measurement is distinct from the standard one: in no stage of the measurement we obtain the eigenvalues of the measured variable. Each system in the small ensemble contributes the shift of the pointer proportional not to one of the eigenvalues, but to the expectation value. This essential novel point has been repeatedly stressed by the inventors of protective measurement (see, e.g. Aharonov, Anandan and Vaidman 1996). As we know, in the orthodox interpretation of quantum mechanics, the expectation values of variables are not considered as physical properties of a single system, as only one of the eigenvalues is observed in the outcome of the standard measuring procedure and the expectation value can only be defined as a statistical average of the eigenvalues. However, for protective measurements, we obtain the expectation value directly for a single system and not as a statistical average of eigenvalues for an ensemble. Since the expectation value of a variable can be directly measured for a single system, it must be a physical characteristic of a single system, not of an ensemble (e.g. as a statistical average of eigenvalues). This is a definite conclusion we can reach by the analysis of protective measurement.

In the following we will show that the mass and charge density can be measured by protective measurement as expectation values of certain variable for a single quantum system, and thus it is the physical property of the system (Aharonov and Vaidman 1993). Consider again a quantum system in a discrete nondegenerate energy eigenstate $\psi(x)$. The interaction Hamiltonian for measuring the value of an observable $A_n$ in the state assumes the same form as

---

[5] Note that the spreading of the wave packet of the pointer also puts a limit on the time of the interaction (Dass and Qureshi 1999).

[6] It can be argued that only observables that commute with the system's Hamiltonian can be protectively measured with absolute certainty for a single system (Rovelli 1994; Uffink 1999).

Eq. (3):

$$H_I = g(t)PA_n \qquad (5)$$

where $A_n$ is a normalized projection operator on small regions $V_n$ having volume $v_n$, which can be written as follows:

$$A_n = \begin{cases} \dfrac{1}{v_n}, & x \in V_n \\ 0, & x \notin V_n \end{cases} \qquad (6)$$

Then a protective measurement of $A_n$ will yield the following result:

$$\langle A_n \rangle = \frac{1}{v_n} \int_{v_n} |\psi(x)|^2 dv = |\psi_n|^2 \qquad (7)$$

It is the average of the density $|\psi(x)|^2$ over the small region $V_n$. When $v_n \to 0$ and after performing measurements in sufficiently many regions $V_n$ we can find the whole density distribution $|\psi(x)|^2$. For a charged system with charge $Q$ the density $|\psi(x)|^2$ times the charge yields the effective charge density $Q|\psi(x)|^2$. In particular, an appropriate adiabatic measurement of the Gauss flux out of a certain region will yield the value of the total charge inside this region, namely the integral of the effective charge density $Q|\psi(x)|^2$ over this region (Aharonov and Vaidman 1993; Aharonov, Anandan and Vaidman 1996). Similarly, we can measure the effective mass density of the system in principle by an appropriate adiabatic measurement of the flux of its gravitational field. Therefore, protective measurement shows that the mass and charge of a single quantum system described by the wave function $\psi(x)$ is indeed distributed throughout space with effective mass density $m|\psi(x)|^2$ and effective charge density $Q|\psi(x)|^2$ respectively. For instance, in the double-slit experiment of an electron, a protective measurement of the charge density will show that there is a charge of $e/2$ in each of the slits when the electron is passing the slits.

Although protective measurement strongly suggests a realistic interpretation of the wave function, it does not directly tell us what the wave function is. It may describe a physical wave or field, as suggested by the inventors of protective measurement (Aharonov and Vaidman 1993)[7]. It is also possible that the wave function describes some kind of ergodic motion of particles, though this view was rejected by Aharonov and Vaidman (1993). Correspondingly, the mass and charge density may result from a physical field or the ergodic motion of a particle. These two explanations are essentially different in that a field exists throughout space simultaneously, whereas the ergodic motion of a particle exists throughout space in an essentially local way. As we will see in the next section, they can be tested by further analyzing the mass and charge density distributions of a quantum system, and the former has already been refuted by experimental observations.

---

[7] Note that protective measurement itself does not entail the field explanation, and it just shows that there is some sort of density distributing in space. The mass and charge density may result from a physical field or the ergodic motion of a particle. As we think, it seems that the existence of some observables such as position in quantum mechanics already suggests the particle explanation. A field has no position property. Thus the expectation value of a variable must be a physical characteristic of the motion of a particle, not that of a field.

## 4. Why the wave function is not a physical field

Now we will investigate the implication of the existence of mass and charge density for the field explanation of the wave function. For the sake of simplicity, we will restrict our discussions to the wave function of a single quantum system. The conclusion can be readily extended to many-body systems[8].

If the wave function is a physical field, then its mass and charge density will simultaneously distribute in space. This has two disaster consequences at least. One is that charge will not be quantized; the total charge inside a very small region can be much smaller than an elementary charge for a single quantum system. This obviously contradicts the common expectation that charge should be quantized. But maybe our expectation needs to be revised. So this consequence is not fatal for the field explanation of the wave function. The other is that the wave function will not satisfy the superposition principle. For example, for the wave function of a single electron, different spatial parts of the wave function will have gravitational and electrostatic interactions, as these parts have mass and charge *simultaneously*.

Let's analyze the second consequence in more detail. Interestingly, the so-called Schrödinger-Newton equation, which was proposed for other purposes (Diosi 1984; Penrose 1998), just describes the gravitational self-interaction of the wave function. The equation for a single quantum system can be written as

$$i\hbar \frac{\partial \psi(x,t)}{\partial t} = -\frac{\hbar^2}{2m}\nabla^2 \psi(x,t) - Gm^2 \int \frac{|\psi(x',t)|^2}{|x-x'|} d^3x' \psi(x,t) + V\psi(x,t) \qquad (8)$$

where $m$ is the mass of the quantum system, $V$ is an external potential, and $G$ is Newton's gravitational constant. Much work has been done to study the mathematical properties of this interesting equation (see, e.g. Harrison, Moroz and Tod 2003; Moroz, Penrose and Tod 1998; Moroz and Tod 1999; Salzman 2005). Some experimental schemes have been also proposed to test its physical validity (Salzman and Carlip 2006). As we will see, although such gravitational self-interactions cannot yet be excluded by experiments[9], the existence of electrostatic

---

[8] It has been argued that for many-body systems the wave functions living on configuration space can hardly be considered as real physical fields (see, e.g. Monton 2002, 2006). However, this objection is not conclusive, and one can still insist on the reality of the wave function living on configuration space by some metaphysical arguments (see, e.g. Albert 1996; Lewis 2004; Wallace and Timpson 2009). Different from this objection, I will in this section propose a more serious objection to the field interpretation, according to which even for a single quantum system the wave function living in real space cannot be taken as a physical field either. Moreover, the reason is not metaphysical but physical, i.e. that the field interpretation contradicts both quantum mechanics and experimental observations.

[9] It has been argued that the existence of a self-interaction term in the Schrödinger-Newton equation does not have a consistent Born rule interpretation (Adler 2007). The reason is that the probability of simultaneously finding a

self-interaction already contradicts experimental observations.

If there is also an electrostatic self-interaction, then the equation for a free quantum system with mass $m$ and charge $Q$ will be

$$i\hbar \frac{\partial \psi(x,t)}{\partial t} = -\frac{\hbar^2}{2m}\nabla^2 \psi(x,t) + (kQ^2 - Gm^2)\int \frac{|\psi(x',t)|^2}{|x-x'|} d^3x' \psi(x,t) \quad (9)$$

where $k$ is the Coulomb constant. Note that the gravitational self-interaction is an attractive force, while the electrostatic self-interaction is a repulsive force. It has been shown that the measure of the potential strength of a gravitational self-interaction is $\varepsilon^2 = \left(\frac{4Gm^2}{\hbar c}\right)^2$ for a free particle with mass $m$ (Salzman 2005). This quantity represents the strength of the influence of self-interaction on the normal evolution of the wave function; when $\varepsilon^2 \approx 1$ the influence will be significant. Similarly, for a free charged particle with charge $Q$, the measure of the potential strength of the electrostatic self-interaction is $\varepsilon^2 = \left(\frac{4kQ^2}{\hbar c}\right)^2$. As a typical example, for a free electron with charge $e$, the potential strength of the electrostatic self-interaction will be $\varepsilon^2 = \left(\frac{4ke^2}{\hbar c}\right)^2 \approx 1\times 10^{-3}$. This indicates that the electrostatic self-interaction will have significant influence on the evolution of the wave function of a free electron. If such an interaction indeed exists, it should have been detected by precise experiments on charged microscopic particles. As another example, consider the electron in the hydrogen atom. Since the potential of its electrostatic self-interaction is of the same order as the Coulomb potential produced by the nucleus, the energy levels of hydrogen atoms will be significantly different from those predicted by quantum mechanics and confirmed by experimental observations. Therefore, the electrostatic self-interaction cannot exist for the wave function of a charged quantum system. Since the field explanation of the wave function entails the existence of such electrostatic self-interactions, it cannot be right, i.e. the wave function cannot be a description of a physical field.

One may object to the above argument with the example of classical electromagnetic field. Electromagnetic field is a field, but it has no self-interaction. Thus a field does not require the existence of self-interaction. However, this is a common misunderstanding. The crux of the matter is that the non-existence of electromagnetic self-interaction results from the fact that electromagnetic field itself has no charge. If the electromagnetic field had charge, then there would also exist electromagnetic self-interaction due to the nature of field, namely the simultaneous existence of its properties in space. In fact, although electromagnetic field has no

---

particle in different positions is zero. However, in a realistic interpretation of quantum mechanics where the wave function is regarded as a real physical entity rather than merely as a probability amplitude, the existence of gravitational self-interaction term seems quite natural. For example, the field interpretation can be consistent with conventional quantum measurement via a dynamical collapse process. As we think, one convincing objection is that if there is a self-gravitational interaction for the wave function of a charged particle, then there will also exist an electrostatic self-interaction because the charge density always accompanies the mass density, while the existence of electrostatic self-interaction is already inconsistent with experimental observations (see below). If this objection is valid, then the Schrödinger-Newton equation will be wrong, and moreover, the approach of semiclassical gravity will also be excluded (cf. Salzman and Carlip 2006).

electromagnetic self-interaction, it does have gravitational self-interaction; the simultaneous existence of energy densities in different spatial locations for an electromagnetic field must generate a gravitational interaction, though the interaction is too weak to be detected by current technology.

One may further object that the superposition principle in quantum mechanics already prohibits the existence of the above self-interactions. But this is just the key point we use to argue against the field explanation of the wave function. Let's state the argument more explicitly. If the wave function of a charged quantum system is a physical field, then the different spatial parts of this field will have gravitational and electrostatic interactions. But the superposition principle in quantum mechanics, which has been verified within astonishing precision, does not permit the existence of the remarkable electrostatic self-interaction[10]. Therefore, the field explanation of the wave function is already refuted by the superposition principle of quantum mechanics.

## 5. Why classical ergodic models fail

If the wave function is not a description of a physical field, then exactly what does the wave function describe? This naturally leads us to the second view that takes the wave function as a description of some sort of ergodic motion of particles. On this view, the effective mass and charge density are formed by time average of the motion of a charged particle, and they distribute in different locations at different moments. At every instant, there is only a localized particle with mass and charge. Thus there will not exist any self-interaction for the wave function, and this view can be consistent with quantum mechanics and experimental observations. In fact, if the mass and charge density does not exist in different regions simultaneously as the field explanation holds, they can only be formed by the ergodic motion of a particle. As a result, the wave function must be a description of some sort of ergodic motion of particles.

There are indeed some realistic interpretations of quantum mechanics that attempt to explain the wave function in terms of the ergodic motion of particles. A well-known example is the stochastic interpretation of quantum mechanics (e.g. Nelson 1966). Nelson (1966) derived the Schrödinger equation from Newtonian mechanics via the hypothesis that every particle of mass $m$ is subject to a Brownian motion with diffusion coefficient $\hbar/2m$ and no friction. In more technical terms, the quantum mechanical process is claimed to be equivalent to a classical Markovian diffusion process. On this interpretation, particles have continuous trajectories but no velocities, and the wave function is a statistical average description of their ergodic motion.

However, it has been pointed out that the classical stochastic interpretations are inconsistent with quantum mechanics (Glabert, Hänggi and Talkner 1979; Wallstrom 1994)[11]. Glabert, Hänggi and Talkner (1979) argued that the Schrödinger equation is not equivalent to a Markovian process, and the various correlation functions used in quantum mechanics do not have the properties of the correlations of a classical stochastic process. Wallstrom (1994) further showed that one must add by hand a quantization condition, as in the old quantum theory, in order to recover the Schrödinger equation, and thus the Schrödinger equation and the Madelung hydrodynamic equations are not equivalent. In fact, Nelson (2005) also showed that there is an empirical difference between the predictions of quantum mechanics and his stochastic mechanics when considering quantum entanglement and nonlocality. For example, for two widely-separated but entangled harmonic oscillators, the two theories predict totally different statistics; stochastic mechanics predicts that measurements of the position of the first one at time $T$ (oscillation period) and the position of the second one at time 0 do not interfere with each other, while quantum mechanics predicts that there exists a strong correlation between them.

---

[10] Note that the superposition principle may be violated when considering gravity (see, e.g. Penrose 1996). But even if there is such a violation, its cause is probably not the self-gravitational interaction. The reason is that if there is a self-gravitational interaction for the wave function of a charged quantum system, then there will also exist an electrostatic self-interaction because the charge always accompanies the mass, while the existence of electrostatic self-interaction is already inconsistent with experimental observations.

[11] Note that some variants of stochastic interpretation assume that the motion of particles is discrete random jump (Bell 1986; Vink 1993; Barrett 2005). But since each random jump is generally limited in a local region, and in particular, it reduces to the Bohmian trajectory in the continuum limit (Vink 1993), these models cannot be consistent with quantum mechanics either. For example, they cannot explain the existence of effective mass and charge density measureable by protective measurement, which is proportional to the square of the absolute value of the wave function. For a more detailed discussion see the last section.

In addition, it can be generally argued that the classical ergodic models that assume continuous motion of particles cannot be consistent with quantum mechanics (Aharonov and Vaidman 1993; Gao 2010). In order to see this let's examine whether the continuous motion of particles can generate the charge density $Q|\psi(x,t)|^2$ at the level of time average. Consider an electron in a one-dimensional box in an energy eigenstate such as the first excited state (Aharonov and Vaidman 1993). Its wave function has a node at the center of the box, where its charge density is zero. The electron performs a very fast motion in the box. At a particular time the charge density is either zero (if the electron is not there) or singular (if the electron is inside the infinitesimally small region including the space point in question). But during a time interval, the motion of the electron will generate a charge density cloud with an effective charge density. The question is whether this density can assume the same form as $e|\psi(x)|^2$. Since the effective charge density is proportional to the amount of time the electron spends in a given position, the electron must be in the left half of the box half of the time and in the right half of the box half of the time. But it can spend no time at the center of the box where the effective charge density is zero; in other words, it must move at infinite velocity at the center. Certainly, the appearance of infinite velocity or velocity faster than light may be not a fatal problem, as our discussion is entirely in the context of non-relativistic quantum mechanics and especially the infinite potential assumed in the example is also an ideal situation. However, it is hard to understand why the electron speeds up at the node and where the infinite energy required for the acceleration comes from. Moreover, the sudden acceleration of the electron near the node will result in large radiation (Aharonov, Anandan and Vaidman 1993), which is inconsistent with both the predictions of quantum mechanics and experimental observations. Maybe one can also assume in an *ad hoc* way that the accelerating electron does not radiate here in order to make the model be consistent with quantum mechanics and experimental observations.

Let's further consider a superposition of two energy eigenstate respectively limited in two widely-separated boxes. In this example, even if one assumes that the electron can move with an infinite velocity (e.g. at the nodes), it cannot move from one box to another due to the limitation of box walls. Therefore, any sort of continuous motion cannot generate the charge density $e|\psi(x)|^2$ at the level of time average. One may still object that this is merely an artificial result of the idealization of infinite potential. But even in this ideal example, the model should also be able to generate the charge density by means of some sort of ergodic motion of the electron. In fact, there is a very similar situation in double-slit experiment. The wave function of a single electron passes through two channels that are well separated in space. The wave function disappears outside the channels for all practical purposes, and the electron can only move inside the channels (otherwise the electron will be detected outside the channels, which contradicts experimental observations). Again, a classical ergodic model cannot explain this experiment.

As we think, there is a general objection to all classical ergodic models. Any classical ergodic model will inevitably introduce a finite ergodic time parameter, which is needed to generate the effective mass and charge density, because it must take a finite time for the particle to continuously move throughout all regions where the wave function is not zero. However, it can be argued that no finite time scale is permitted to exist for the ergodic motion. First of all, the existence of a finite time scale, denoted by $T_c$, is inconsistent with the standard quantum theory, as there is no such a time constant in the theory. Next, if there exists a time scale $T_c$, then when the measuring time $T$ of protective measurement is shorter than $T_c$ (i.e. $T < T_c$), the measurement result will be not the expectation value of a variable such as charge density, as no whole time average can be obtained. This also contradicts the prediction of protective measurement. As an extreme example, consider a spatial superposition state $\psi_L + \psi_R$, where $\psi_L$ and $\psi_R$ are Gaussian wave packets and their centers are well separated in space. When $T < T_c$, the particle has no enough time to move throughout the whole regions including both $L$ and $R$. Then the result of a protective position measurement will be not the expectation value of $\psi_L + \psi_R$, but the eigenvalue corresponding to $\psi_L$ or $\psi_R$. Moreover, the results distribution is also different from that predicted by protective measurement. When $T < T_c$, the distribution of position measurement results will concentrate near $L$ and $R$, while according to protective measurement, the distribution should concentrate near the midpoint between $L$ and $R$. In fact, for protective measurement, during any period of time the pointer of the measuring device always shifts by an amount proportional to the expectation value of the measured variable, rather than to one of its eigenvalues. Thus the expectation value can be associated with any short period of time. Certainly, pointer shifts during short time intervals are practically unobservable since they are much smaller than the uncertainty, and only when the total shift accumulated during the whole period of measurement is much larger than the width of the initial distribution it becomes observable (Aharonov and Vaidman 1993).

Therefore, we conclude that the continuous ergodic motion of particles cannot generate the effective mass and charge density measurable by protective measurement, and the classical ergodic models cannot be consistent with quantum mechanics. As a result, the wave function cannot be a description of the continuous ergodic motion of particles.

## 6. The wave function as a description of quantum motion of particles

The failure of classical ergodic models does not exclude all possible ergodic motion of particles. In this section, we will argue that another different kind of motion – random discontinuous motion can naturally generate the effective mass and charge density measurable by protective measurement, and what the wave function describes is probably such quantum motion of particles, which is essentially discontinuous and random (Gao 1993, 1999, 2000, 2003, 2006a, 2006b, 2010).

If the motion of a particle is not continuous but discontinuous and random, and the probability density of the particle being in certain positions is proportional to the square of the absolute value of its wave function at every instant, then the particle can readily move throughout all possible regions where the wave function is nonzero during an arbitrarily short time interval near a given instant. This will solve the problems plagued by the classical ergodic models. The discontinuous ergodic motion requires no existence of a finite time scale. A particle undergoing discontinuous motion can also move from one region to another spatially separated region, no matter whether there is an infinite potential wall between them. Besides, discontinuous motion can also solve the problems of infinite velocity and accelerating radiation. The reason is that no classical velocity and acceleration can be defined for discontinuous motion, and energy and momentum will require new definition and new understanding as in quantum mechanics. Thus it seems that the discontinuous ergodic motion of particles can in principle generate the effective mass and charge density measurable by protective measurement, and thus the wave function is probably a description of such random discontinuous motion of particles.

In some sense, the above interpretation of the wave function seems to be an inevitable consequence of protective measurement. According to protective measurement, a charged quantum system has effective mass and charge density distributing in space, proportional to the square of the absolute value of its wave function. There are two possible ways to explain the existence of the mass and charge density; one is to take the wave function as a description of some kind of physical field, and the mass and charge density of this field exists throughout space simultaneously, the other is to take the wave function as a description of some sort of ergodic motion of particles, and the effective mass and charge density formed by such motion exists throughout space in an essentially local way. The first view has been rejected because it entails the existence of a remarkable electrostatic self-interaction that contradicts experimental observations. Thus the wave function can only be a description of some sort of ergodic motion of particles. Since the classical ergodic models have also been excluded, the ergodic motion of particles cannot be continuous and must be essentially discontinuous. Besides, when considering the randomness of the results of conventional quantum measurement, such motion must be also random. Therefore, what the wave function describes can only be random discontinuous motion of particles.

In fact, by assuming the wave function is a (complete) description for the motion of particles, we can reach the random discontinuous motion in a more direct way, independent of the analysis of protective measurement. If the wave function $\psi(x,t)$ is a description of the state of motion for a single particle, then the quantity $|\psi(x,t)|^2 dx$ not only gives the probability of the particle being found in an infinitesimal space interval $dx$ near position $x$ at instant $t$ (as in standard quantum mechanics), but also gives the objective probability of the particle being there[12]. This accords with the well-accepted assumption that the probability distribution of the measurement outcomes of a property is the same as the actual distribution of the property in the measured state. Then at instant $t$ the particle may appear in any location where the probability density $|\psi(x,t)|^2$ is nonzero, and during an infinitesimal time interval near instant $t$, the particle will move throughout the whole space where the wave function $\psi(x,t)$ spreads[13], though it is still in one position at each instant. Moreover, the density distribution of its positions is equal to the probability density $|\psi(x,t)|^2$. Obviously, this kind of motion is essentially random and discontinuous.

---

[12] It has been argued that the probability related to the wave function should be the objective character of a quantum system, and not merely the display of measurement results (see, e.g. Popper 1959; Bunge 1973; Shimony

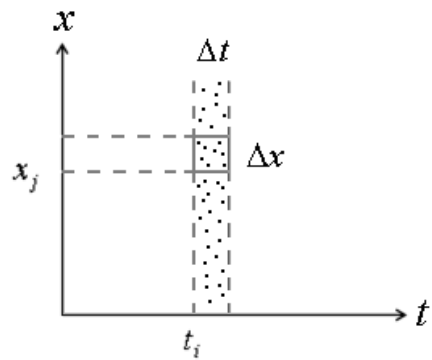

**Fig. 1** The description of RDM of a single particle

The strict mathematical description of random discontinuous motion (RDM henceforth) can be obtained by using the measure theory (see, e.g. Nielsen 1994). Consider the motion state of a single particle in finite intervals $\Delta t$ and $\Delta x$ near a space-time point $(t_i, x_j)$ as shown in Fig. 1. For RDM, the position of the particle forms a random discontinuous point set in the whole space for the time interval $\Delta t$ near the instant $t_i$[14]. Accordingly, there is a local discontinuous point

---

1993). According to Bunge (1973), the probability of an event is an objective property inherent in things; likewise a probability distribution is also an objective property of a physical system. For example, the probability of a transition from one state of a system to another state is just as objective as the speed of the transition.

[13] Since an infinitesimal time interval near a given instant contains infinitely many instants, all possible positions can be distributed in this time interval.

[14] A random discontinuous point set can be defined as a set of points $(t,x)$ in continuous space and time, for which the function $x(t)$ is discontinuous and random at all instants. The definition of a discontinuous function is as follows. Suppose $A$ is an open set in $\Re$ (say an interval $A=(a,b)$, or $A=\Re$), and $f:A\to\Re$ is a function. Then $f$ is discontinuous at $x\in A$, if $f$ is not continuous at $x$. Note that a function $f:A\to\Re$ is continuous if and only if for every $x\in A$ and every real number $\varepsilon>0$, there

set in the space interval $\Delta x$ near the position $x_j$. The local discontinuous point set represents the motion state of the particle in the finite intervals $\Delta t$ and $\Delta x$ near the space-time point $(t_i, x_j)$. We study its projection in the *t*-axis, namely the corresponding dense instant set in the time interval $\Delta t$. Let *W* be the discontinuous trajectory or world-set of the particle and *Q* be the square region $[x_j, x_j + \Delta x] \times [t_i, t_i + \Delta t]$. The dense instant set can be denoted by $\pi_t(W \cap Q) \subset \Re$, where $\pi_t$ is the projection on the *t*-axis. According to the measure theory, we can define the Lebesgue measure:

$$M_{\Delta x, \Delta t}(x_j, t_i) = \int_{\pi_t(W \cap Q) \subset \Re} dt \tag{10}$$

Since the sum of the measures of all such dense instant sets in the time interval $\Delta t$ is equal to the length of the continuous time interval $\Delta t$, we have:

$$\sum_j M_{\Delta x, \Delta t}(x_j, t_i) = \Delta t \tag{11}$$

Then we can define the measure density:

$$\rho(x,t) = \lim_{\substack{\Delta x \to 0 \\ \Delta t \to 0}} M_{\Delta x, \Delta t}(x,t) / (\Delta x \cdot \Delta t) \tag{12}$$

The limit exists for a random discontinuous point set. This provides a strict mathematical description of the point distribution situation for the above local discontinuous point set. We call this measure density position measure density.

Since the local discontinuous point set represents the motion state of the particle, the position measure density $\rho(x,t)$ will be a descriptive quantity for the RDM of a single particle. It represents the relative frequency of the particle appearing in an infinitesimal space interval $dx$ near position $x$ during an infinitesimal interval $dt$ near instant $t$. From Eq. (12) we can see that $\rho(x,t)$ satisfies the normalization relation, namely $\int_{-\infty}^{+\infty} \rho(x,t)dx = 1$. Furthermore, we can define the position measure flux density $j(x,t)$ through the relation $j(x,t) = \rho(x,t)v(x,t)$, where $v(x,t)$ is the velocity of the local discontinuous point set. It describes the change of the position measure density with time. Due to the conservation of measure, $\rho(x,t)$ and $j(x,t)$ satisfy the following equation:

$$\frac{\partial \rho(x,t)}{\partial t} + \frac{\partial j(x,t)}{\partial x} = 0 \tag{13}$$

---

exists a real number $\delta > 0$ such that whenever a point $z \in A$ has distance less than $\delta$ to $x$, the point $f(z) \in \Re$ has distance less than $\varepsilon$ to $f(x)$.

The position measure density $\rho(x,t)$ and the position measure flux density $j(x,t)$ provide a complete description for the RDM of a single particle.

It is very natural to extend the description of the motion of a single particle to the motion of many particles. For the RDM state of *N* particles, we can define a joint position measure density $\rho(x_1, x_2, ... x_N, t)$. This represents the relative probability of the situation in which particle 1 is in position $x_1$, particle 2 is in position $x_2$, ... , and particle *N* is in position $x_N$. In a similar way, we can define the joint position measure flux density $j(x_1, x_2, ... x_N, t)$. It satisfies the joint measure conservation equation:

$$\frac{\partial \rho(x_1, x_2, ... x_N, t)}{\partial t} + \sum_{i=1}^{N} \frac{\partial j(x_1, x_2, ... x_N, t)}{\partial x_i} = 0 \qquad (14)$$

When these *N* particles are independent, the joint position measure density $\rho(x_1, x_2, ... x_N, t)$ can be reduced to the direct product of the position measure density of each particle, namely $\rho(x_1, x_2, ... x_N, t) = \prod_{i=1}^{N} \rho(x_i, t)$. It is worth noting that the joint position measure density $\rho(x_1, x_2, ... x_N, t)$ and the joint position measure flux density $j(x_1, x_2, ... x_N, t)$ are not defined in the three-dimensional real space, but defined in the 3N-dimensional configuration space. As we will see later, these two quantities can further constitute the wave function, and the many-body wave functions thus defined also live on the 3N-dimensional configuration space.

With respect to the RDM of a particle, the motion of the particle is completely discontinuous and random. The probability density for the particle to appear at position $x$ at instant $t$ is $\rho(x,t)$. There is no evolution law for the position state of a particle, and the trajectory function $x(t)$ is random and discontinuous at every instant. However, the discontinuity of RDM is absorbed into the motion state of a particle, which is defined during an infinitesimal time interval, by the descriptive quantities of position measure density $\rho(x,t)$ and position measure flux density $j(x,t)$. Therefore, the evolution law for the motion state of a particle will contain no discontinuities and can be a continuous equation.

By assuming that the nonrelativistic evolution equation of RDM is the Schrödinger equation[15], the wave function $\psi(x,t)$ can be uniquely expressed by the position measure density $\rho(x,t)$ and the position measure flux density $j(x,t)$:

---

[15] Note that Gao (2006b) also gave a heuristic derivation of the Schrödinger equation in terms of RDM.

$$\psi(x,t) = \sqrt{\rho(x,t)} e^{iS(x,t)/\hbar} \quad (15)$$

where $S(x,t) = m \int_{-\infty}^{x} \frac{j(x',t)}{\rho(x',t)} dx'$. Since $\rho(x,t)$ and $j(x,t)$ provide a complete description of the RDM of a single particle, the wave function $\psi(x,t)$ also provides a complete description of the RDM of a single particle[16].

The new interpretation of the wave function in terms of RDM of particles can be taken as a natural realistic alternative to the orthodox view. According to the standard probability interpretation of the wave function, the square of the absolute value of a N-particle wave function, which can be denoted by $|\psi(x_1, x_2, ... x_N, t)|^2 dx_1 dx_2 ... dx_N$, gives the probability of particle 1 being found in the infinitesimal interval $dx_1$ near position $x_1$ and particle 2 being found in the infinitesimal interval $dx_2$ near position $x_2$, ... , and particle N being found in the infinitesimal interval $dx_N$ near position $x_N$. By contrast, according to the new interpretation, the square of the absolute value of the wave function not only gives the probability of a particle *being found* in certain locations, but also gives the objective probability of the particle *being* there. For example, $|\psi(x_1, x_2, ... x_N, t)|^2 dx_1 dx_2 ... dx_N$ also represents the objective probability of particle 1 being in the infinitesimal interval $dx_1$ near position $x_1$ and particle 2 being in the infinitesimal interval $dx_2$ near position $x_2$, ... , and particle N being in the infinitesimal interval $dx_N$ near position $x_N$. Certainly, the transition process from "being" to "being found", which is closely related to the notorious quantum measurement problem, also needs to be explained. We will discuss this issue in the next section.

## 7. Further discussions

If the wave function is really a description of the ergodic motion of particles, which is random and discontinuous in nature, then the main realistic interpretations of quantum mechanics will be either rejected or revised. In this last section, we will further discuss the implications of the suggested picture of RDM for the interpretation of quantum mechanics.

To begin with, the de Broglie-Bohm theory will be problematic[17]. The theory takes the wave

---

[16] In some sense, the wave function can also be regarded as a guiding agent of the RDM of particles when considering the instantaneous change of position. For example, $\rho(x,t) = |\psi(x,t)|^2$ can be regarded as an instantaneous intrinsic property of an particle, which determines the probabilities of the particle appearing in an infinitesimal spatial interval $dx$ near position $x$ at instant $t$.

[17] It is worth noting that some objections have already been raised concerning the reality of Bohmian particles in terms of weak measurement and protective measurement (Englert, Scully, Süssmann and Walther 1992; Aharonov

function as a physical field (i.e. Ψ-field) and further adds the non-ergodic motion of Bohmian particles to interpret quantum mechanics. This is obviously inconsistent with the suggested new interpretation of the wave function. Concretely speaking, taking the wave function as a Ψ-field will lead to the existence of electrostatic self-interaction that contradicts both quantum mechanics and experimental observations. Moreover, inasmuch as the wave function has charge density distribution in space for a charged quantum system, there will also exist an electromagnetic interaction between it and the Bohmian particles. This is inconsistent with the predictions of quantum mechanics and experimental observations either.

Certainly, one can eliminate the electromagnetic interaction between the Ψ-field and Bohmian particles by depriving the Bohmian particles of mass and charge. But they will be not real particles any more. Then in what sense the de Broglie-Bohm theory provides a realistic interpretation of quantum mechanics? One may also want to deprive the Ψ-field of mass and charge density to eliminate the electrostatic self-interaction. But, on the one hand, the theory will break its physical connection with quantum mechanics, as the wave function in quantum mechanics has mass and charge density according to our analysis, and on the other hand, since protective measurement can measure the mass and charge density for a single quantum system, the theory will be unable to explain the measurement results either. Although de Broglie-Bohm theory can still exist in this way as a mathematical tool for experimental predictions (somewhat like the orthodox interpretation it tries to replace), it obviously departs from the initial expectations of de Broglie and Bohm, and as we think, it already fails as a physical theory because of losing its explanation ability.

Next, the ontology of the many-worlds interpretation and dynamical collapse theories needs to be revised from field to particle. The wave function is not a field but a description of the ergodic motion of particles. However, we may still have ontology-revised many-worlds interpretation and dynamical collapse theories[18]. The left problem is to determine which is basically right: the former denies the existence of wavefunction collapse while the latter admit its existence.

As we think, it can be further argued that there is only one world and quantum mechanics is also a one-world theory in terms of RDM. The key point is that quantum superposition exists in a form of time division by means of the RDM of particles, and there is only one observer (as well as one quantum system and one measuring device) all along in a continuous time flow during quantum evolution. For an observer in a quantum superposition the brain state of the observer changes discontinuously, while for a classical observer in classical mechanics his brain state evolves continuously. There is no essential difference between these two situations. For both situations the brain state of the observer is always definite at each instant, and the states at different instants can be different. If there is only one world in classical mechanics, so does in quantum mechanics.

In addition, if the above quantum superposition indeed corresponds to many observers in

---

and Vaidman 1996; Aharonov, Englert and Scully 1999; Aharonov, Erez and Scully 2004). But it seems that these objections can be answered by noticing that protective measurement is in fact a way of measuring the effect of the Ψ-field rather than that of the Bohmian particle (see, e.g. Drezet 2006). Anyway these analyses indicates that the motion of Bohmian particle is not ergodic, and the time averages of Bohmian particle's positions typically differ markedly from the ensemble averages (Aharonov, Erez and Scully 2004).

[18] Note that the ontology-revised many-worlds interpretation in terms of RDM seems very similar to Bell (1981)'s Everett (?) theory. However, as we will argue, this theory is in fact improper because RDM implies that there is only one world and quantum mechanics is also a one-world theory (cf. Barrett, Leifer and Tumulka 2005).

many worlds, then each observer can only exist in a discontinuous dense instant set, a time sub-flow of the continuous time flow. As a result, at every instant only one of these observers exists, and all other observers do not exist at all. This poses another serious objection to the many worlds theory. Anyway, RDM universally exist for all systems including microscopic particles, measuring devices and observers. For a microscopic particle in a quantum superposition, there is only one particle being in an indefinite state, and there are no many particles each of which is in a definite state in one of the many worlds. Then for a measuring device or an observer in a quantum superposition, the conclusion should be the same, no matter what conscious experience the observer in a quantum superposition has[19].

Therefore, we conclude that there is only one world all along during quantum evolution in the framework of RDM. As a result, the many-worlds interpretation will be also problematic. Moreover, our definite conscious experience and the definite measurement outcomes (e.g. positions of pointer) in the unique world will further demand that there exists an objective process of wavefunction collapse, which is responsible for the transition from microscopic uncertainty to macroscopic (approximate) certainty (e.g. in Schrödinger's cat thought experiment). Thus the dynamical collapse theories will be in the right direction by admitting wavefunction collapse. However, their ontology should be revised from field to particle, and certainly, the physical origin of the wavefunction collapse also needs to be found. It has been argued that the discreteness of spacetime may inevitably result in the collapse of the wave function[20], and the compete evolution law of RDM in discrete spacetime will naturally include the dynamical collapse of the wave function. In particular, the inherent random motion of particles just provides the random source to collapse the wave function (Gao 2006a, 2006b). But more study is still needed before we can finally solve the quantum measurement problem (e.g. preferred basis problem) and completely understand the meaning of quantum theory.

## Acknowledgments

I am very grateful to Dean Rickles, Huw Price, Hans Westman, Antony Valentini, and Lev Vaidman for helpful discussions. I am also grateful to the participants of the 2nd PIAF Workshop on Foundations for discussions. This work was supported by the Postgraduate Scholarship in Quantum Foundations provided by the Unit for History and Philosophy of Science and Centre for Time (SOPHI) of the University of Sydney.

## References

Adler, S. L. (2007). Comments on proposed gravitational modifications of Schrödinger dynamics and their experimental implications. *J. Phys.* A 40, 755-764.
Aharonov, Y., Anandan, J. and Vaidman, L. (1993). Meaning of the wave function, *Phys. Rev.* A 47, 4616.

---

[19] It can be guessed that the conscious experience of an observer in a quantum superposition of two different perceptions is probably not one of the definite perceptions in the superposition. This conjecture can be tested by experiments in principle.

[20] For other interesting suggestions see, e.g. Feynman (1995) and Penrose (1996).

Aharonov, Y., Anandan, J. and Vaidman, L. (1996). The meaning of protective measurements, *Found. Phys.* 26, 117.

Aharonov, Y., Englert, B. G. and Scully M. O. (1999). Protective measurements and Bohm trajectories, *Phys. Lett.* A 263, 137.

Aharonov, Y., Erez, N. and Scully M. O. (2004). Time and Ensemble Averages in Bohmian Mechanics. *Physica Scripta* 69, 81–83.

Aharonov, Y. and Vaidman, L. (1993). Measurement of the Schrödinger wave of a single particle, *Phys. Lett.* A 178, 38.

Aharonov, Y. and Vaidman, L. (1996). About position measurements which do not show the Bohmian particle position, in *Bohmian Mechanics and Quantum Theory: An Appraisal*, J. T. Cushing, A. Fine, S. Goldstein, eds. Dordrecht: Kluwer Academic.

Albert, D. (1996). Elementary Quantum Metaphysics, in James Cushing, Arthur Fine, and Sheldon Goldstein (eds.), *Bohmian Mechanics and Quantum Theory: An Appraisal*. Dordrecht: Kluwer, 277–284.

Barrett, J., Leifer, M. and Tumulka, R. (2005). Bell's jump process in discrete time. *Europhys. Lett.* 72, 685.

Bell, J. S. (1981) Quantum mechanics for cosmologists, in C. Isham, R. Penrose and D. Sciama eds, *Quantum Gravity 2*. Oxford: Clarendon Press. pp. 611-637.

Bell, J. S. (1986). Beables for quantum field theory. *Phys. Rep.* 137, 49-54.

Bell, J. (1990) Against 'measurement', in A. I. Miller (ed.), *Sixty-Two Years of Uncertainty: Historical Philosophical and Physics Enquiries into the Foundations of Quantum Mechanics*. Berlin: Springer. pp. 17-33.

Bohm, D. (1952). A suggested interpretation of quantum theory in terms of "hidden" variables, I and II. *Phys. Rev.* 85, 166-193.

Bohm D. and Hiley, B.J. (1993) *The Undivided Universe: An Ontological Interpretation of Quantum Theory*. London: Routledge.

Brown, H. R., Dewdney, C. and Horton, G. (1995) Bohm particles and their detection in the light of neutron interferometry. *Found. Phys.* 25, 329.

Dass, N. D. H. and Qureshi, T. (1999). Critique of protective measurements. *Phys. Rev.* A 59, 2590.

de Broglie, L. (1928). in: *Electrons et photons: Rapports et discussions du cinquième Conseil de Physique Solvay*, eds. J. Bordet. Paris: Gauthier-Villars. pp.105. English translation: G. Bacciagaluppi and A. Valentini (2009), *Quantum Theory at the Crossroads: Reconsidering the 1927 Solvay Conference*. Cambridge: Cambridge University Press.

Diósi, L. (1984). Gravitation and the quantum-mechanical localization of macro-objects. *Phys. Lett.* A 105, 199-202.

Drezet, A. (2006). Comment on "Protective measurements and Bohm trajectories", *Phys. Lett.* A 350, 416.

Dürr, D., Goldstein, S., and Zanghì, N. (1997). "Bohmian mechanics and the meaning of the wave function", in Cohen, R. S., Horne, M., and Stachel, J., eds., Experimental Metaphysics — Quantum Mechanical Studies for Abner Shimony, Volume One; *Boston Studies in the Philosophy of Science* 193, Boston: Kluwer Academic Publishers.

Englert, B. G., Scully, M. O., Süssmann, G., Walther, H. (1992). *Z. Naturforsch*. Surrealistic Bohm Trajectories. 47a, 1175.

Everett, H. (1957). 'Relative state' formulation of quantum mechanics. *Rev. Mod. Phys.* 29, 454-462.

Feynman, R. (1995). *Feynman Lectures on Gravitation*. B. Hatfield (ed.), Reading, Massachusetts: Addison-Wesley.

Gao, S. (1993). A suggested interpretation of quantum mechanics in terms of discontinuous motion (unpublished manuscript).

Gao, S. (1999). The interpretation of quantum mechanics (I) and (II). physics/9907001, physics/9907002.

Gao, S. (2000). *Quantum Motion and Superluminal Communication*, Beijing: Chinese Broadcasting & Television Publishing House. (in Chinese)

Gao, S. (2003). *Quantum: A Historical and Logical Journey.* Beijing: Tsinghua University Press. (in Chinese)

Gao, S. (2006a). A model of wavefunction collapse in discrete space-time. *Int. J. Theor. Phys.* 45(10), 1943-1957.

Gao, S. (2006b). *Quantum Motion: Unveiling the Mysterious Quantum World.* Bury St Edmunds, Suffolk U.K.: Arima Publishing.

Gao, S. (2010). Meaning of the wave function. arXiv:1001.5085 [physics.gen-ph].

Ghirardi, G. C., Grassi, R. and Benatti, F. (1995). Describing the macroscopic world: Closing the circle within the dynamical reduction program. *Found. Phys.*, 25, 313–328.

Goldstein, S. and Teufel, S. (2001). "Quantum spacetime without observers: Ontological clarity and the conceptual foundations of quantum gravity", in Callender, C. and Huggett, N., eds., *Physics meets Philosophy at the Planck Scale*, Cambridge: Cambridge University Press.

Grabert, H., Hänggi, P. and Talkner, P. (1979). Is quantum mechanics equivalent to a classical stochastic process? *Phys. Rev. A* 19, 2440–2445.

Harrison, R., Moroz, I. and Tod, K. P. (2003). A numerical study of the Schrödinger-Newton equations. *Nonlinearity* 16, 101-122.

Holland, P. (1993). *The Quantum Theory of Motion: An Account of the de Broglie-Bohm Causal Interpretation of Quantum Mechanics*. Cambridge: Cambridge University Press.

Lewis, P. (2004). Life in Configuration Space, *British Journal for the Philosophy of Science* 55, 713–729.

Monton, B. (2002). Wave Function Ontology. *Synthese* 130, 265-277.

Monton, B. (2006). Quantum mechanics and 3N-dimensional space. *Philosophy of Science*, 73(5), 778–789.

Moroz, I. M., Penrose, R. and Tod, P. (1998). Spherically-symmetric solutions of the Schrödinger-Newton equations. *Class. Quant. Grav.* 15, 2733.

Moroz, I. M. and Tod, K. P. (1999). An analytical approach to the Schrödinger-Newton equations. *Nonlinearity* 12, 201-16.

Nelson, E. (1966). Derivation of the Schrödinger equation from Newtonian mechanics. *Phys. Rev.* 150, 1079–1085.

Nelson, E. (2005). The mystery of stochastic mechanics, manuscript 2005-11-22.

Penrose, R. (1996). On gravity's role in quantum state reduction. *Gen. Rel. Grav.* 28, 581.

Penrose, R. (1998). Quantum computation, entanglement and state reduction. *Phil. Trans. R. Soc. Lond. A* 356, 1927.

Rovelli, C. (1994). Meaning of the wave function - Comment, *Phys. Rev. A* 50, 2788.

Salzman, P. J. (2005). Investigation of the Time Dependent Schrödinger-Newton Equation, Ph.D. Dissertation, University of California at Davis.

Salzman, P. J. and Carlip, S. (2006). A possible experimental test of quantized gravity. arXiv: gr-qc/0606120.

Uffink, J. (1999). How to protect the interpretation of the wave function against protective measurements, *Phys. Rev.* A 60, 3474.

Valentini, A. (1997). On Galilean and Lorentz invariance in pilot-wave dynamics. *Phys. Lett. A* 228, 215–222.

Valentini, A. (2009). The nature of the wave function in de Broglie's pilot-wave theory. Talk in PIAF '09 New Perspectives on the Quantum State Conference. http://pirsa.org/09090094/.

Vink, J. C. (1993). Quantum mechanics in terms of discrete beables. *Phys. Rev. A* 48, 1808.

Wallace, D. and Timpson, C. G. (2009). Quantum mechanics on spacetime I: spacetime state realism. PhilSci Archive 4621. Forthcoming in the British Journal for the Philosophy of Science.

Wallstrom, T. (1994). Inequivalence between the Schrödinger equation and the Madelung hydrodynamic equations. *Phys. Rev.* A 49, 1613–1617.